%% file: main.tex
\documentclass[sigconf]{acmart}

\usepackage{siunitx}
\usepackage{nicefrac}
\usepackage{cleveref}
\usepackage{tikz}
\usepackage{pgfplots}

\usetikzlibrary{decorations.pathreplacing, angles, quotes, arrows, positioning, backgrounds, shapes, decorations.fractals, spy}
\pgfplotsset{compat=newest}
\usepgfplotslibrary{groupplots}
\usepgfplotslibrary{dateplot}
\usetikzlibrary{patterns}
\tikzset{font={\small}}
\pgfdeclarelayer{background}
\pgfdeclarelayer{foreground}
\definecolor{lssblue}{rgb}{0.121,0.231,0.4}
\definecolor{lssred}{rgb}{0.76,0.3,0.19}

\newcommand{\pystencils}{{\em pystencils}}
\newcommand{\lbmpy}{{{\em lbmpy}}}
\newcommand{\walberla}{\textsc{waLBerla}}
\newcommand{\sympy}{{{\em SymPy}}}

\newcommand{\lagoon}{LAGOON}
\newcommand{\scalable}{SCALABLE}

\definecolor{color1}{rgb}{0.00392156862745098, 0.45098039215686275, 0.6980392156862745}
\definecolor{color2}{rgb}{0.8705882352941177, 0.5607843137254902, 0.0196078431372549}
\definecolor{color3}{rgb}{0.00784313725490196, 0.6196078431372549, 0.45098039215686275}
\definecolor{color4}{rgb}{0.8352941176470589, 0.3686274509803922, 0.0}

\date{February 2023}
\title{Scalable Flow Simulations with the Lattice Boltzmann Method}
\subtitle{Invited Paper}

\begin{document}
\input{author.tex}

\input{abstract.tex}

\maketitle

\section{Introduction}
\input{introduction.tex}

\section{What is the LBM}
\label{sec:lbm}
\input{lbm.tex}

\section{The LBM framework \walberla{}}
\label{sec:walberla}
\input{walberla.tex}

\section{LaBS}
\label{sec:labs}
\input{labs.tex}

\section{Benchmark cases}
\label{sec:benchmarks}
\input{benchmarks}

\subsection{Turbulent Channel}
\label{subsec:benchmarks:turbulent}
\input{turbulent-channel}

\subsection{Lagoon}
\label{subsec:benchmarks:lagoon}
\input{lagoon.tex}

\section{Conclusions}
\label{sec:conclusions}
\input{conclusion.tex}

\section{Acknowledgement}
\label{sec:acknowledgements}
\input{acknowledgement.tex}

\bibliographystyle{acm}
\bibliography{references}
\end{document}

%% file: author.tex
\author{Markus Holzer}
\email{markus.holzer@cerfacs.fr}
\orcid{0000-0001-5499-5857}
\author{Gabriel Staffelbach}
\email{gabriel.staffelbach@cerfacs.fr}
\orcid{0000-0001-5499-5857}
\author{Ilan Rocchi}
\email{rocchi@cerfacs.fr}
\affiliation{%
  \institution{CERFACS}
  \city{Toulouse}
  \country{France}
}

\author{Jayesh Badwaik}
\email{j.badwaik@fz-juelich.de}
\orcid{0000-0002-5252-8179}
\author{Andreas Herten}
\email{a.herten@fz-juelich.de}
\orcid{0000-0002-7150-2505}
\affiliation{%
  \institution{J{\"u}lich Supercomputing Center}
  \city{J{\"u}lich}
  \country{Germany}
}

\author{Radim Vavrik}
\email{radim.vavrik@vsb.cz}

\author{Ondrej Vysocky}
\email{ondrej.vysocky@vsb.cz}
\orcid{0000-0001-7849-2744}

\author{Lubomir Riha}
\email{lubomir.riha@vsb.cz}
\orcid{0000-0002-1017-5766}
\affiliation{%
    \institution{IT4Innovations, VSB - TU Ostrava}
    \city{Ostrava}
    \country{Czech Republic}
}

\author{Romain Cuidard}
\email{romain.cuidard@csgroup.fr}
\affiliation{%
    \institution{CS GROUP}
    \city{Paris}
    \country{France}
}

\author{Ulrich Ruede}
\email{ulrich.ruede@fau.de}
\orcid{0000-0001-8796-8599}
\affiliation{%
    \institution{Friedrich Alexander Universt{\"a}t Erlangen-N{\"u}rnberg and CERFACS}
    \country{Erlangen, Germany and Toulouse, France}
}

\keywords{Lattice-Boltzmann, Supercomputing, Massively-Parallel, Computational
Fluid Dynamics, Lagoon Landing Gear, GPUs}
\copyrightyear{2023}
\acmYear{2023}
\setcopyright{licensedothergov}\acmConference[CF '23]{20th ACM International
Conference on Computing Frontiers}{May 9--11, 2023}{Bologna, Italy}
\acmBooktitle{20th ACM International Conference on Computing Frontiers (CF
'23), May 9--11, 2023, Bologna, Italy}
\acmPrice{15.00}
\acmDOI{10.1145/3587135.3592176}
\acmISBN{979-8-4007-0140-5/23/05}


\renewcommand{\shortauthors}{Jayesh Badwaik et al.}

\ccsdesc[500]{Computing methodologies~Massively parallel and high-performance simulations}

%% file: abstract.tex
\begin{abstract}
    The primary goal of the EuroHPC JU project \scalable{} is to develop an industrial Lattice Boltzmann Method (LBM)-based computational fluid dynamics (CFD) solver capable of exploiting current and future extreme scale architectures, expanding current capabilities of existing industrial LBM solvers by at least two orders of magnitude in terms of processor cores and lattice cells, while preserving its accessibility from both the end-user and software developer's point of view. This is accomplished by transferring technology and knowledge between an academic code (\walberla{}) and an industrial code (\textsc{LaBS}). This paper briefly introduces the characteristics and main features of both software packages involved in the process. We also highlight some of the performance achievements in scales of up to tens of thousand of cores presented on one academic and one industrial benchmark case.
\end{abstract}

%% file: introduction.tex
In the \scalable{} project, eminent industrial and academic partners teamed up to achieve the leap to extreme scale performance, scalability, and energy efficiency of an industrial Computational Fluid Dynamics (CFD) software based on the Lattice Boltzmann Method (LBM). 
LBMs have already evolved to become trustworthy alternatives to conventional CFD and in several engineering applications they are shown to be roughly an order of magnitude faster than Navier-Stokes approaches in a fair comparison and in comparable scenarios. 
In the context of EuroHPC~\footnote{\url{https://eurohpc-ju.europa.eu/}}, the LBM is especially well-suited to exploit advanced supercomputer architectures through vectorization, accelerators, and massive parallelization.

\textbf{\walberla{}}
In the public domain research \walberla{} code, superb performance and unlimited scalability have been demonstrated, 
reaching more than a trillion ($10^{12}$) lattice cells already on petascale systems.
\walberla{} performance excels because of its uncompromising unique, architecture-specific automatic generation of optimized compute kernels, 
together with carefully designed parallel data structures and communication routines. \walberla{}, however, is primarily used in academic cases and targeted toward experts in the field of High Performance Computing (HPC).

\textbf{LaBS}
The industrial \textsc{LaBS} CFD software package is focused on industrial use cases and, thus, it already has proven capabilities to simulate such problems at a high level of maturity. Performance-wise, however, it still shows room for improvement. Therefore, \scalable{} will transfer the leading-edge performance technology from the \walberla{} framework to \textsc{LaBS}, thus breaking the silos between the scientific computing world and physical flow modeling world to deliver improved efficiency and scalability for \textsc{LaBS} to be prepared for the upcoming European Exascale systems.

In this paper, we describe the characteristics of both software packages and highlight some of achievements in the project. We briefly introduce the Lattice Boltzmann method in \Cref{sec:lbm}. Then, we describe the two software packages waLBerla and LaBS respectively in \Cref{sec:walberla} and \Cref{sec:labs}. In \Cref{sec:benchmarks}, we describe the optimizations carried out as a part of the project and their result on the performance of a couple of benchmark test cases. We finally describe our conclusions in \Cref{sec:conclusions}.

%% file: lbm.tex
Over the past two decades, the LBM has grown to become a reliable alternative to classical CFD 
that is based on solving the discretized Navier-Stokes equations. Its advantages are a simple basic algorithmic structure 
and very low computational cost per degree of freedom. 

The basic idea of the LBM is to solve the lattice Boltzmann equation given by:
\begin{align}
\label{eq:LBCollision}
  f_i^* \left(\boldsymbol{x}, t\right) &= f_i^{\mathrm{eq}} \left(\boldsymbol{x}, t\right) + \left(1 - \Omega \right) f_i^{\mathrm{neq}} \left(\boldsymbol{x}, t\right) \\
  \label{eq:LBStreaming}
 f_i \left(\boldsymbol{x} + \boldsymbol{c}_i \Delta t, t + \Delta t\right) &= f_i^* \left(\boldsymbol{x}, t\right),
\end{align}
where $f_i(\boldsymbol{x}, t)$ is the Particle Distribution Function (PDF) at position $\boldsymbol{x}$ and time $t$ for a discrete particle with velocity $\boldsymbol{c}_i$, $\Delta t$ is the time step, $\boldsymbol{f}^{\mathrm{eq}} \left(\boldsymbol{x}, t\right)$ and $\boldsymbol{f}^{\mathrm{neq}} \left(\boldsymbol{x}, t\right)$ are the equilibrium and non-equilibrium PDFs, which depends on the local fluid properties. Furthermore, $\Omega$ is the non-linear collision operator that is entirely local while the streaming step shown in \cref{eq:LBStreaming} is linear and typically requires only first neighborhood memory accesses \cite{lbm_book}.

Due to these algorithmic properties, the LBM has emerged as a popular alternative to the traditional Navier-Stokes-based approach. Thus, a scientific consensus has emerged that rigorously optimized LBM implementations can offer approximately ten times faster simulations than their Navier-Stokes-based counterparts for certain classes of applications. However, this performance advantage is only observed when comparing the two approaches in a fair scenario, where the same data is retrieved and the same sampling rate is used (thus using explicit schemes in both cases) \cite{lbm_book}.

The computational efficiency of the LBM has made it particularly attractive for industries with computationally demanding simulations, such as the aeronautics and automotive industries. These industries rely heavily on CFD simulations to optimize their designs and improve their products' performance. With its ability to provide fast and accurate simulations, LBM has found its way into industrial usage and is expected to become even more relevant in the future~\cite{lagoon3}.

One significant advantage of the LBM due to its high locality is the ability to exploit massively parallel machines, making it particularly suitable for high-performance computing environments. This capability has been demonstrated not only on x86-type CPUs but also on heterogeneous systems, including GPUs \cite{Holzer2021, Bauer2021lbmpy}. Recent advancements in LBM implementation on these platforms are expected to lead to significant improvements in computational efficiency and the ability to tackle larger problems. This makes the LBM a natural hot candidate in the algorithmic choices for the Exascale era~\cite{LBMExascale}. Furthermore, the reduced turnaround time and computational costs enhance the energy efficiency of modern industrial design activities~\cite{lagoon3}.

In summary, the LBM's computational efficiency, scalability, and algorithmic specificity make it a valuable tool for simulating fluid dynamics, particularly in industries with computationally demanding simulations. With ongoing improvements in LBM implementations and high-performance computing capabilities, it is expected to continue to grow in importance in the years to come.

%% file: walberla.tex
\walberla{} is a modern open-source multiphysics software framework specifically designed to address the performance challenge in computational science and engineering that is implemented in C++ (version C++17). Especially simulation of physically relevant phenomena requires billions of grid points to ensure sufficient domain resolution due to the different scales of the problem. This typically demands the usage of large-scale HPC systems \cite{rettinger_2022, BAUER2021478, LBMExascale}. The architecture of the framework is designed in a modular way to assure productivity, reusability, and maintainability for future development. Furthermore, to guarantee the scalability of the software only fully distributed data structures are supported. Thus each process only holds information about itself and neighboring processes, but no global information is communicated. The distribution of the work is done via the Message Passing Interface (MPI) on the inter-node level and via MPI, OpenMP, or using accelerators in the form of GPUs on the intra-node level \cite{Holzer2021}.

\begin{figure}[htb]
	\centering
	\begin{tikzpicture}[font=\sffamily]
	\fill[color1] (0,0) rectangle ++(2,2);
	\fill[color2] (0,2) rectangle ++(2,2);
	\fill[color3] (6,0) rectangle ++(2,2);
	\fill[color4] (6,2) rectangle ++(2,2);
	\fill[color1] (2,0) rectangle ++(1,1);
	\fill[color1] (2,1) rectangle ++(1,1);
	\fill[color1] (2,2) rectangle ++(1,1);
 
	\fill[color2] (2,3) rectangle ++(1,1);
	\fill[color2] (3,3) rectangle ++(1,1);
	\fill[color2] (4,3) rectangle ++(1,1);
 
	\fill[color4] (4,2) rectangle ++(1,1);
	\fill[color4] (5,2) rectangle ++(1,1);
	\fill[color4] (5,3) rectangle ++(1,1);

    \fill[color3] (4,0) rectangle ++(1,1);
    \fill[color3] (5,0) rectangle ++(1,1);
    \fill[color3] (4,1) rectangle ++(1,1);
	\fill[color3] (5,1) rectangle ++(1,1);
	\fill[color1] (3,0) rectangle ++(0.5,0.5);
	\fill[color1] (3,0.5) rectangle ++(0.5,0.5);
	\fill[color1] (3.5,0) rectangle ++(0.5,0.5);
	\fill[color2] (3,2) rectangle ++(0.5,0.5);
	\fill[color2] (3,2.5) rectangle ++(0.5,0.5);
	\fill[color2] (3.5,2.5) rectangle ++(0.5,0.5);
	\fill[color3] (3,1) rectangle ++(0.5,0.5);
	\fill[color3] (3.5,1) rectangle ++(0.5,0.5);
	\fill[color3] (3.5,0.5) rectangle ++(0.5,0.5);
	\fill[color4] (3.5,1.5) rectangle ++(0.5,0.5);
	\fill[color4] (3,1.5) rectangle ++(0.5,0.5);
	\fill[color4] (3.5,2) rectangle ++(0.5,0.5);

    \draw[step=.1,black,thin] (6,0) grid (8,2);
    \draw[step=.05,black,thin] (5,0) grid (6,1);

	\draw[step=2,black,thick] (0,0) grid (8,4);
	\draw[step=1,black,thick] (2,0) grid (6,4);
	\draw[step=0.5,black,thick] (3,0) grid (4,3);
	
    \node[right] at (0, -0.5) {Processes:};
 
	\fill[color1] (2, -0.75) rectangle ++(0.5,0.5);
	\node[right] at (2.5, -0.5) {$P_1$};
	
	\fill[color2] (3.5, -0.75) rectangle ++(0.5,0.5);
	\node[right] at (4,-0.5) {$P_2$};
	
	\fill[color3] (5,-0.75) rectangle ++(0.5,0.5);
	\node[right] at (5.5,-0.5) {$P_3$};
	
	\fill[color4] (6.5,-0.75) rectangle ++(0.5,0.5);
	\node[right] at (7,-0.5) {$P_4$};
	
	\end{tikzpicture}
    \caption{Illustration of the domain partitioning in \walberla{}. First, the domain bounding box is subdivided into cuboidal subdomains. These blocks can then be refined in a 2:1 ratio. Afterward, the load balancing step and thus the assignment to the processes happen (here the \emph{block forest} is distributed to four processes). Each of these then allocates a fixed number of cells for each block as a linear array. For illustration purposes, this is done for two blocks on the third process with a block size of twenty cells in each direction. More information about \walberla{}'s structure can be found in \cite{BAUER2021478}.}
    \label{fig:walberla}
\end{figure}
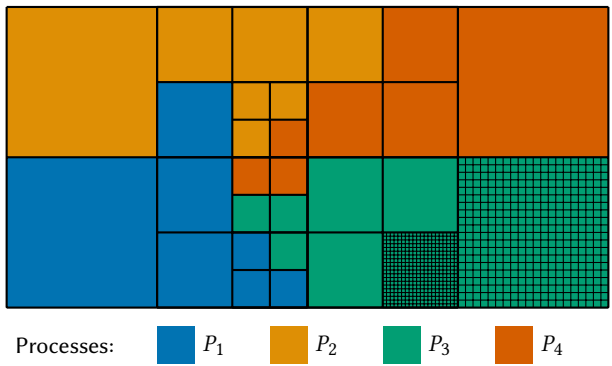

In order to run massively parallel simulations on hundreds of thousands of processes efficiently, not only the main timestep needs to be optimized but also the domain setup, the load balancing, and post-processing routines need to be addressed. In \walberla{} this is done by decomposing the domain in a block-structured grid. Each of these blocks contains a fixed number of cells and thus a small regular subdomain of the problem. A software design like this provides several advantages. First of all, a separation of data allocation and domain partitioning can be achieved since the blocks can be allocated first via a proxy class not containing any actual cells. This proxy structure is managed in an octree structure, namely \walberla{}'s \emph{block forest}. This results in the advantage that the domain setup and the load-balancing become orders of magnitude simpler because the octree is orders of magnitude smaller than it would be using the actual cells directly at the beginning. Finally, this results also in higher flexibility during the simulation, which makes it possible to use highly complex simulation techniques like adaptively refined meshes at massive scales \cite{walberla2013, waLBerla_python, schornbaum2016, schornbaum2018}. An illustration of the domain decomposition in blocks is shown in \cref{fig:walberla}.

\begin{figure}
\begin{tikzpicture}[framed,background rectangle/.style={fill=black!10, rounded corners, draw=black!90, dashed}]
\def\scale{1.3};
\def\w{3};
\def\lw{.85mm};
\tikzset{
	state/.style={
		rectangle split ,
		rectangle split parts=2,
		rectangle split horizontal,
		rectangle split draw splits=false,
		rounded corners,
		draw=black, very thick,
		minimum height=4em,
		text width=2.55cm,
		inner sep=2pt,
		text centered,
	}
}

\tikzset{
	state2/.style={
		rectangle split ,
		rectangle split parts=3,
		rectangle split horizontal,
		rectangle split draw splits=false,
		rounded corners,
		draw=black, very thick,
		minimum height=4em,
		text width=1.64cm,
		inner sep=2pt,
		text centered,
	}
}

\tikzset{
	state3/.style={
		draw=black,very thick,
		rounded corners =.3cm,
		minimum height=4em,
		text width=2cm,
		text centered,
	}
}
\tikzset{
	state4/.style={
		rounded corners,
		draw=black, very thick,
		minimum height=4em,
		text width=5.2cm,
		text centered,
	}
}

	\node (a) [state4, left color=color1, right color=color1] at (0,10) {\textbf{lbmpy /  pystencils}};
	\node (b) [state3, fill=color1] at (-3.9, 10) {Model creation};
	
	\node (c) [state3, fill=color2, below=of b] {Code generation and optimisation};
	\node (d) [state2, left color=color2, right color=color2, below=of a] {\nodepart{one} Compute kernel \nodepart{two} Boundary conditions
	\nodepart{three} Packing, Unpacking, Interpolation};
	
	\node (e) [state3, fill=color3, below=of c] {Backends};
	\node (f) [state4, left color=color3, right color=color3, below=of d] {CPU (C-Code) and GPU (CUDA)};

	\node (g) [state3, fill=color4, below=of e] {Execution};
	\node (h) [state, left color=color4, right color=color4, below=of f] {\nodepart{one} Interactively with IPython \nodepart{two} MPI distributed with \walberla{}};

        \draw[-stealth, line width= \lw, color1]  (0, 9.3) to[out=270, in=90]  (-2,8.35);
	\draw[-stealth, line width= \lw, color1]  (0, 9.3) to[out=270, in=90]  (0,8.35);
	\draw[-stealth, line width= \lw, color1]  (0, 9.3) to[out=270, in=90]  (2,8.35);

	\draw[-stealth, line width= \lw, color2]  (-2, 6.98) to[out=270, in=90]  (0,6.04);
	\draw[-stealth, line width= \lw, color2]  (0, 6.98) to[out=270, in=90]  (0,6.04);
	\draw[-stealth, line width= \lw, color2]  (2, 6.98) to[out=270, in=90]  (0,6.04);
	
	\draw[-stealth, line width= \lw, color3]  (0.018, 4.68) -- (-1,3.75);
	\draw[-stealth, line width= \lw, color3]  (-0.018, 4.68) -- (1,3.75);

\end{tikzpicture}
\caption{Complete workflow of combining \lbmpy{} and \walberla{} for MPI parallel execution. Furthermore, \lbmpy{} can be used as a stand-alone package for prototyping.}
\label{fig:walberlaChart}
\end{figure}
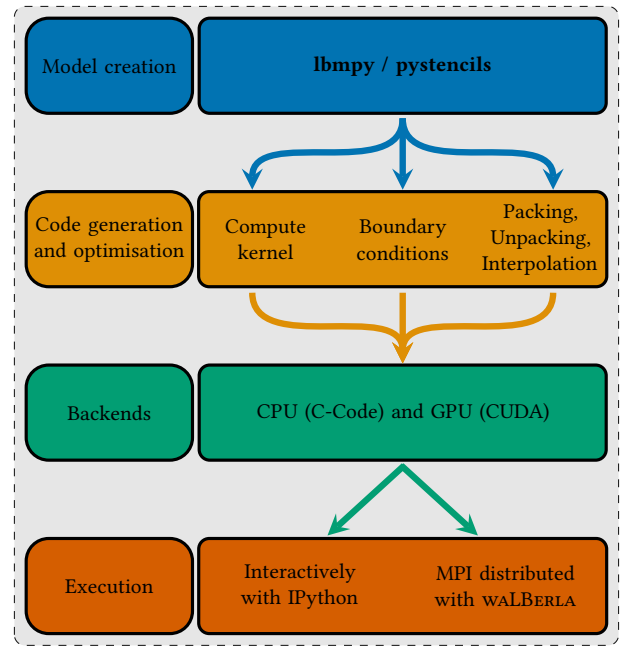

Due to the nature of the LBM, maximal performance on modern massively parallel supercomputers of the compute kernels is limited by the memory bandwidth of a compute node \cite{Bauer2021lbmpy, Holzer2021, OpenLB}. On the other side, its high data locality contributes to great scalability. In order to achieve maximal performance, careful optimizations of the most demanding hot spots of the code (the kernels) and the communication need to be realized. In general, low-level optimizations in C or Fortran language are not very flexible for the development of new models and portable enough for smooth deployment on heterogeneous HPC architectures of various vendors. In \walberla{} this is addressed by using the metaprogramming libraries \pystencils{} and \lbmpy{}, two high-level Python frameworks to automatically generate the software components in an architecture- and problem-specific way \cite{Bauer2019pystencils, Bauer2021lbmpy, Holzer2021}. Their usage also significantly improves the readability and maintainability of the code. An illustration of the combination of \pystencils{} / \lbmpy{} with \walberla{} is shown in \cref{fig:walberlaChart}.

On the top level, a fully symbolic high-level formulation of the LBM is used. That is enabled via the open-source \sympy{} library, which also performs powerful symbolic simplifications of the mathematical expressions \cite{sympy}. Due to these mathematical transformations, highly specialized problem-specific LBM compute kernels can be generated with a minimal number of FLOPs. 

From this symbolic description, we create an Abstract Syntax Tree (AST) in the \pystencils{} Intermitted Representation (IR). 
In this tree representation, we introduce architecture-specific AST nodes for example to represent the loop nest or pointer access in the later kernels. In this representation, spatial access information is encoded in the form of \pystencils{} \emph{fields}. Furthermore, loop-based optimizations like spatial blocking or OpenMP parallelization can be employed or constant evaluations can be moved out of the loop nest to further reduce the computational cost.

Since the LBM compute kernel is defined symbolically with all its field data accesses in symbolic, it is also possible to automatically derive compute kernels for the boundary conditions as well as for packing, unpacking, and interpolation routines for grid transitions. These kernels are used to fill communication buffers for MPI.

Finally, the intermediate representation of the compute, boundary, and packing/unpacking kernels is transformed either by the C or the CUDA backend of \pystencils{}. The task of the backend is to print the AST as a C-function with a clearly defined interface. Each function takes raw pointers for array accesses together with their representative shape and stride information as well as all remaining free parameters. This simple and consistent interface makes it possible for the kernels to be called from a large variety of languages because it is usually easy to call C-functions from most languages. In the case of Python, the Python C-API can be employed. Thus, we can provide a powerful interactive development environment by utilizing \lbmpy{}/\pystencils{} as stand-alone packages where the generated kernels are provided as Python functions via Pythons C-API.

Additionally, this simple low-level interface provided by the \pystencils{} backends makes it possible to combine the highly optimized compute kernels with existing HPC frameworks like the multiphysics \walberla{} framework. In this case, \walberla{} provides the necessary boilerplate code to integrate the kernels nicely in the framework as well as all additional functionality needed to run complex simulations at large scales. This includes a domain decomposition, functionality for complex meshes in the form of STL files, and parallel I/O to enable post-processing for large-scale simulations.

%% file: labs.tex
\textsc{LaBS} is an industrial CFD software used to perform Aeroacoustics, Aerodynamics, and  Aerothermal simulation on complex cases. Currently, it is mainly executed on \num{500} to \num{2000} processors in production. \textsc{LaBS} is a C++ software based on MPI.

One of the strengths of \textsc{LaBS} is the parallel mesh generator which removes the theoretical bottleneck for volume mesh generation during the simulation.  Numerical schemes in \textsc{LaBS} are specified using a declarative (non-imperative) Domain Specific Language (DSL). \textsc{LaBS} parses the DSL to automatically generate the scheme, the associated Halo structures, and the communication routines for the Halo exchange. The size of the halo structures are optimized for the scheme automatically. The DSL can handle complex spatial and temporal features in the scheme, which are used at mesh transitions and near the boundaries.

Currently, there are two official schemes with a different collision model:
\begin{itemize}
  \item D3Q19DRT (Double Relaxation Time)
  \item D3Q19HRR (Hybrid Recursive Regularization)
\end{itemize}

More schemes can be added as a plugin to \textsc{LaBS} through the means of a Physics module. The module takes advantage of the various facilities provided by the \textsc{LaBS} kernel. In particular, it can do:
\begin{itemize}
    \item Precomputation and partitioning of the geometry its pieces on relevant cores.
    \item Automatic parallelization, distributed computation on all available cores, and ensuring automated transfer between cores of the required information for computations.
    \item Output data extraction and files generation, specifying time operators combined with space operators. This possibly includes spatial and temporal sub-sampling to minimize the size of the resulting storage.
\end{itemize}

In order to ease the work of the numerical developer, many optimizations such as load balancing, memory alignment, etc. are done in the core of the solver. However, some other optimizations must be done within the numerical scheme.

%% file: benchmarks.tex
As a set of use cases for the project, two academic and two industrial benchmark cases were selected. 
The academic 
tests have a known simple solution, 
whereas the industrial use cases are representative of the target applications of the end-user partners of the project. 

%% file: turbulent-channel.tex
This test case simulates a turbulent flow between two parallel and infinite plates.
It is one of the simplest turbulent wall-bounded flows. The channel must be sufficiently long to allow the flow to converge toward a solution independent of x. A periodic condition is imposed in the span wise direction. The high friction Reynolds number leads to a fully turbulent flow. 
This test case allows assessing fundamental properties of the physical modeling, like mass conservation and the symmetry and regularity of the solution (especially near the walls). This test case also allows assessing the wall-normal momentum closure (behavior of the shear stresses (laminar, subgrid, and turbulent) along the wall's normal direction).

We use the Direct Numerical Simulation example from \cite{HoyasScaling} at $Re_{\tau}=2000$ can be used as reference example.
This simulation gives $Cf = \frac{\tau_w}{0.5*\rho_{bulk}*U_{bulk}^2} = 0.0041$, with $\tau_w = \rho_w * U_{\tau}^2$.
For a first node above the wall placed at $0.5*\delta_h/20 = \delta_h/40$, the corresponding
$y_+$ is $Re_{\tau}/40 = 50$.

Test case setup:
\begin{itemize}
    \item number of fluid nodes = \num{730615200}
    \item dx\_max = $2.5e-4$
    \item number of iteration =  100
    \item number of refinement level = 2 
\end{itemize}

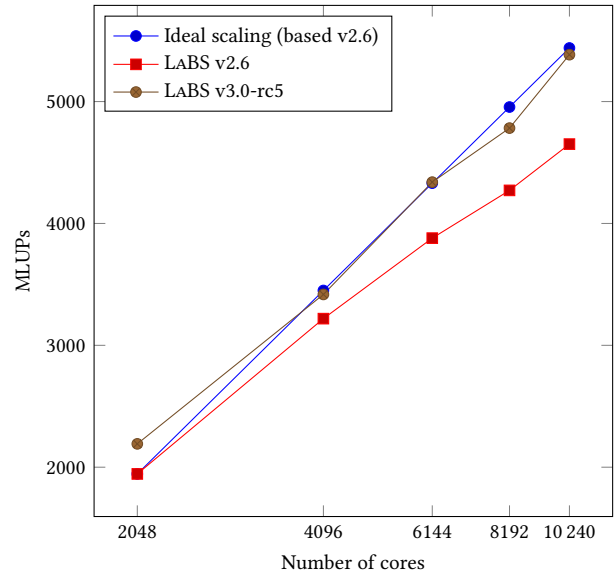
\begin{figure}[h!tp]
\centering
\begin{tikzpicture}
    \begin{axis}[height=0.3\paperheight,width=\linewidth,scaled ticks=false,
    ylabel=MLUPs,
    xlabel=Number of cores,
    xtick={2048,4096,6144,8192,10240},
    xticklabels={\num{2048},\num{4096},\num{6144},\num{8192},\num{10240}},
    yticklabels={\num{1000}, \num{2000}, \num{3000}, \num{4000}, \num{5000}},
    xmode=log,
    ymode=log,
    log basis x={2},
    legend style={at={(0.02,0.98)}, anchor=north west,legend cell align=left}]
    \addplot table [x=Cores, y=$TurbulentChannel_lin$]{data/labs.turbulent.channel.dat};
    \addplot table [x=Cores, y=$TurbulentChannel$]{data/labs.turbulent.channel.dat};
    \addplot table [x=Cores, y=$TurbulentChannel_3.0$]{data/labs.turbulent.channel.dat};
    \legend{Ideal scaling (based v2.6), \textsc{LaBS} v2.6 ,\textsc{LaBS} v3.0-rc5}
    \end{axis}
\end{tikzpicture}

\caption{LaBS executed the turbulent channel strong scaling benchmark on up to 10240 cores. On 10240 cores we
are able to achieve 5000 MLUPs of performance.}
\label{fig:prolb-turbulent-channel}
\end{figure}

The improvement of scalability performance on Turbulent channel between \textsc{LaBS} v2.6 and v3.0-rc5, shown in 
\cref{fig:prolb-turbulent-channel}, is a result of the following three features:
\begin{itemize}
    \item Introduction of the delay feature.
    \item Optimization of PDF data exchanges.
    \item The new collision D3Q19HRR model.
\end{itemize}

The purpose of the delay feature is to make up for the load balancing error. It is similar to the concept of altruism computation. During the solver step, some computations, longer than expected by load balancing, are stopped and delayed, enforcing MPI synchronization. This forced synchronization allows other MPI processes to continue their own computation. Overall, the delay feature reduces the time spent by inactive processes and so improve the scalability of \textsc{LaBS}.

Optimization of PDF exchanges is a feature issued from the analysis of \walberla{}. In the LBM, for computation on standard fluid nodes (do not concern refinement, boundary, moving region, etc.) some directions of the PDF are not used for computation. In that case, \textsc{LaBS} no longer sends those directions to other MPI process. This feature reduces the amount of data transferred between MPI processes, thus increasing the scalability.

In \textsc{LaBS} 3.0, a new collision model named D3Q19HRR is introduced. Although this collision model needs more local computations, it allows removing some filter used for stability in our previous scheme. The consequence in terms of performance is the reduction of the communication stencil. This contributes to an increase in the scalability of the \textsc{LaBS}.

%% file: lagoon.tex
%

\begin{figure}
\centering

\begin{tikzpicture}
  [spy using outlines={rectangle, green, magnification=10,
                       width=2.5cm, height=3cm, connect spies}]
 \node {\includegraphics[width=0.9\linewidth]{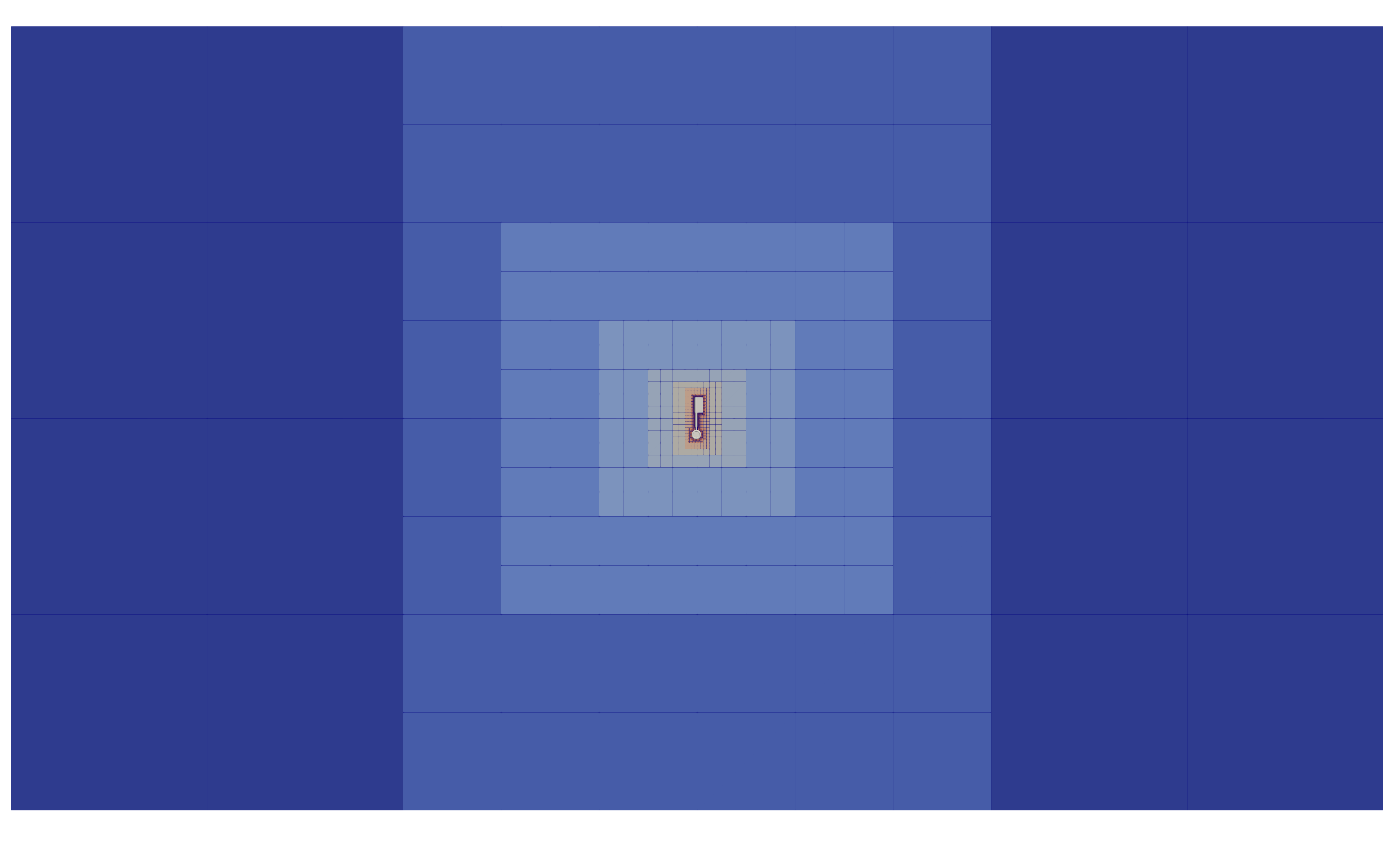}};

  \spy on (-0.015, 0.05) in node [right] at (1.70, -.9);
\end{tikzpicture}

\caption{\lagoon{} setup with \walberla{}. Here we show a cut through the z-aches of the domain setup. The setup shows the block structure used for the simulation. Each of the blocks consists of $24 \times 24 \times 24$ lattice cells. This results in about $1.3 \cdot 10^9$ lattice cells on all grid levels decomposed on \num{94232} blocks (\num{67212} on the finest grid level).}
\label{fig:lagoonSetup}
\end{figure}

To showcase the performance of \walberla{} we use the industrial test case called \lagoon{} (LAnding-Gear nOise database for CAA validatiON), which is a simplified Airbus plane landing gear. The evolution of the flow around the landing gear is studied in a wind tunnel \cite{lagoon1, lagoon2}. 
During the landing phase, the gear is an important source of noise and produces a non-trivial turbulent flow. 
It is convenient for benchmarking CFD codes since the geometry is simple and a large experiment database is available for both aerodynamic and acoustic data. 

The setup for the benchmark case is shown in \cref{fig:lagoonSetup}. It consists of ten mesh refinement levels to achieve a minimal resolution of \num{0.25} millimeters around the landing gear. The simulation of the landing gear with a resolution of \num{0.4} millimeters with the LBM was previously shown here \cite{lagoon3}. In order to avoid any influence of the domain,
we place the landing gear in a $40 \times 20 \times 20$ meters virtual wind tunnel. The wind tunnel consists of an inflow on the left boundary with an inflow velocity of $78.99\,m/s$ and an outflow on the right. With the diameter of the landing gear's wheel $D = 0.3\,m$, we end up with a Reynolds number $\mathrm{Re} = 1.59e^6$  For the remaining boundaries, periodicity conditions are used. The benchmark consists of \num{94232} blocks where \num{67212} blocks form the finest region. Each of the blocks consists of $24 \times 24 \times 24$ lattice cells.

Furthermore, a Multiple-Relaxation-Time (MRT) collision operator is used with a D3Q19 lattice stencil. In order to reduce the allocated memory, we employ the Esoteric Twist pattern with a Structure of Array (SoA) memory layout for the streaming of the distribution functions \cite{Geier2017EsoTwist}. This allows us to work with only a single set of PDFs and further utilize the cache of the hardware. Furthermore, we can run the \lagoon{} benchmark with extremely high resolution even on as little as four nodes, showing the low memory footprint of \walberla{}.
More benchmarks results 
are presented on the project website~\cite{SCALABLEweb}.

\subsubsection{Scalability Results and Analysis}

In order to evaluate the scalability and efficiency of particular performance aspects, the \scalable{} project uses a performance model and analysis methodology developed within the Horizon 2020 POP Center of Excellence~\cite{POPweb}. 

Besides the code performance, also energy-efficiency has been evaluated and optimized using dynamic tuning of the CPU parameters to fit the hardware configuration to the needs of the executed workload. Hardware parameters dynamic tuning and energy consumption measurement are provided by the MERIC runtime system~\cite{meric1}.

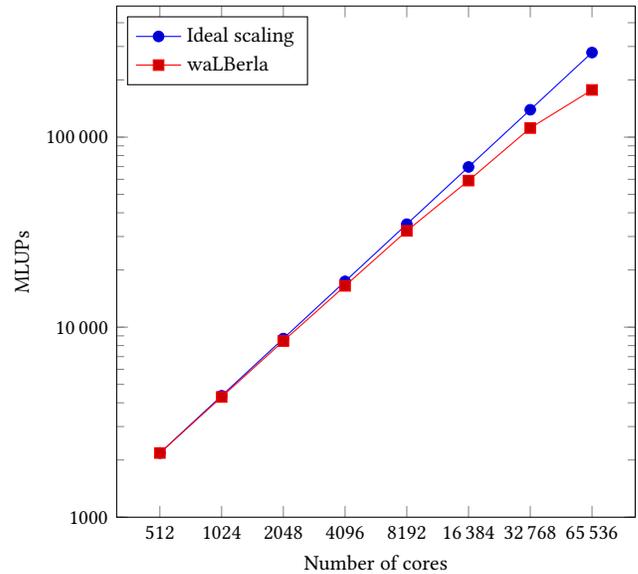
\begin{figure}[h!tp]
\centering
\begin{tikzpicture}
    \begin{axis}[height=0.3\paperheight,width=\linewidth,scaled ticks=false,
    ylabel=MLUPs,
    xlabel=Number of cores,
    xtick={512, 1024, 2048, 4096, 8192, 16384, 32768, 65536},
    xticklabels={512,\num{1024},\num{2048},\num{4096},\num{8192},\num{16384},\num{32768},\num{65536}},
    ymin=1000,
    ytick={1000, 10000, 100000, 1000000},
    yticklabels={\num{1000}, \num{10000}, \num{100000}, \num{1000000}},
    xmode=log,
    ymode=log,
    log basis x={2},
    legend style={at={(0.02,0.98)}, anchor=north west,legend cell align=left}]
    \addplot table [x=Cores, y=$ideal$]{data/walberla.dat};
    \addplot table [x=Cores, y=$Lagoon$]{data/walberla.dat};
    \legend{Ideal scaling, waLBerla}
    \end{axis}
\end{tikzpicture}

\caption{\walberla{} executed \lagoon{} strong scaling benchmark on up to \num{512} compute nodes and \num{65536} cores on the HAWK supercomputer. On \num{65536} we are able to achieve \num{176000}\,MLUPs of performance and a strong scalability of~about 64\,\%. This allows us to update \num{176} billion lattice cells per second.}
\label{fig:waLBerla_StrongScaling}
\end{figure}

In \cref{fig:waLBerla_StrongScaling} the scaling of the \lagoon{} benchmark on up to \num{65536} CPU cores is shown. 
The benchmark was conducted on the HAWK supercomputer \footnote{\url{https://www.hlrs.de/solutions/systems/hpe-apollo-hawk}}. 
HAWK uses two AMD EPYC 7742 CPUs per node where each of the CPUs consists of 64 cores. In order to better classify scaling results, it is important to start looking at single-node performance first. The reason is that scaling alone tells nothing about the actual performance of the code and can even be misleading because slower codes typically allow for better scalability. This can be explained by resource utilization. If shared resources are underused, there are more reserves for parallel execution. However, on the other side, if shared resources are already highly utilized by a sequential program, achieving good scalability becomes significantly more complex. As reported often in literature, efficient LBM implementations are capable of saturating the memory bandwidth of the hardware \cite{Bauer2021lbmpy, Holzer2021, OpenLB}. Thus to determine the theoretical upper limit for the single node performance $P_{\mathrm{max}}$ of an LBM code, we can use:

\begin{align}
\label{eq:roofline}
P_{\mathrm{max}} = \frac{b_\mathrm{s}}{n_\mathrm{b}},
\end{align} 

where $b_\mathrm{s}$ is the machine bandwidth and $n_\mathrm{b}$ is the memory traffic from main memory per cell in bytes. We determine the machine bandwidth by using a STREAM \footnote{\url{https://www.cs.virginia.edu/stream/FTP/Code/stream.c}} benchmark on a single node resulting in a measured maximum bandwidth of $297.9$\,\nicefrac{Gb}{s}. For the D3Q19 lattice stencil used here, we expect $38$ load/stores per cell resulting in $n_\mathrm{b} = 304$\, bytes for double precision computations. Thus, we end up with a theoretical upper limit of $P_{\mathrm{max}} = 977$\,MLUPs. Measuring the performance of just the compute kernel (stream and collide), we end up with \num{874}\,MLUPs, which are satisfying results.

Besides the pure compute kernel, additional components need to be executed to run the \lagoon{} benchmark. This includes computing kernels for boundary handling, interpolation kernels for coarse-to-fine and fine-to-coarse resolution, and communication kernels to exchange data between processes. As pictured on the first data point in \cref{fig:waLBerla_StrongScaling}, adding all the mentioned components still gives us \num{2177}\,MLUPs on \num{512} cores. Thus, we still have a performance of \num{544}\,MLUPs per node. With these results, we can confidently evaluate the strong scaling experiment.

We ran the benchmark starting with 
\num{512} cores for a constant domain size 
to explore the strong scalability limits. 
On \num{65536} cores, we still achieve about 64\,\% performance per core as compared to \num{512} cores. 
On this configuration, the code can carry out an update (i.e. time step) of \num{176} billion lattice cells per second, equivalent to a total performance of \num{176000}\,MLUPs for this complex test case.

%% file: conclusion.tex
In this paper, we have introduced the \scalable{} project and its main objective,
which is to bring together developers from academia and industry and to create an LBM-based CFD solver that is highly efficient and scalable.
We have provided a brief overview of the two software packages that are being examined in this project and presented some of the most recent scaling results.
\walberla{} demonstrates almost perfect scaling up to 65 thousand cores even when tested on industrial cases. The \textsc{LaBS} software has demonstrated a 40\,\% performance improvement between the base and the latest development versions when running on more than 10 thousand cores.
The results presented in this paper are part of a broader set of goals and objectives that were established for the project.
One of these goals was to conduct a performance analysis of both codes in order to reveal their characteristics and bottlenecks. It also helped to identify some of the optimisations in \walberla{} that were applied in LaBS.
Another objective was to successfully adapt \walberla{} to compute complex industrial benchmark cases, which ultimately contributed to the performance and scaling achievements that have been made possible. 
By providing code generation techniques for highly optimized compute kernels, \walberla{} can be used to create building blocks that can be effectively integrated into the \textsc{LaBS} code to maximize its performance potential. 
In addition to performance improvement, significant energy savings were achieved by dynamic tuning of hardware parameters for both codes.

All the development is done using the leading-edge HPC architectures and leads towards efficient utilization of pre-exascale and exascale systems. The vast majority of these systems are empowered by GPU accelerators. Therefore significant effort within the \scalable{} project is currently expended on the efficient GPU acceleration of the compute kernels and inter-GPU communication.

Besides the development and the optimizations aiming at the performance of the LBM solvers on usual x86 CPU and accelerator architectures, we are currently also focusing on wiping out bottlenecks and overcoming obstacles during the deployment and tuning on alternative ARM-based systems that promise high energy efficiency while preserving high performance.


%% file: acknowledgement.tex
This work was supported by the SCALABLE project. This project has received funding from the European High-Performance Computing Joint Undertaking (JU) under grant agreement No 956000. The JU receives support from the European Union’s Horizon 2020 research and innovation programme and France, Germany, and the Czech Republic. 

This work was supported by the Ministry of Education, Youth and Sports of the Czech Republic through the e-INFRA CZ (ID:90140). 

The simulations were performed using the HPE Apollo Hawk national supercomputer at the High Performance Computing Center Stuttgart (HLRS) under grant number TN17/44103.

The authors gratefully acknowledge the Gauss Centre for Supercomputing e.V. \url{www.gauss-centre.eu} for funding this project by providing computing time through the John von Neumann Institute for Computing (NIC) on the GCS Supercomputer JUWELS at Jülich Supercomputing Centre (JSC).